\documentclass{svjour3}                     % onecolumn (standard format)
\smartqed  % flush right qed marks, e.g. at end of proof
\usepackage{graphicx}
\usepackage{hyperref}
\journalname{Quantum Information Processing}
\begin{document}

\title{Temperley-Lieb Algebra, Yang-Baxterization and universal Gate\thanks{This work was supported by NSF of China
(Grant No. 10875026).} }


\author{Gangcheng Wang  \and
        Kang Xue \footnote{The Corresponding Author}\and Chunfang Sun \and Chengcheng Zhou \and Taotao Hu \and Qingyong Wang}


\institute{Gangcheng Wang\at
              Department of Physics, Northeast
Normal University, Changchun, Jilin 130024, P.R.China \\
           \and
           Kang Xue(wanggc@139.com) \at
              Department of Physics, Northeast
Normal University, Changchun, Jilin 130024, P.R.China \\
\email{wanggc@139.com or Xuekang@nenu.edu.cn}\and Chunfang Sun
\at
              Department of Physics, Northeast
Normal University, Changchun, Jilin 130024, P.R.China \\
\and Chengcheng Zhou \at
              Department of Physics, Northeast
Normal University, Changchun, Jilin 130024, P.R.China \\
\and Taotao Hu \at
              Department of Physics, Northeast
Normal University, Changchun, Jilin 130024, P.R.China \\
\and Qingyong Wang \at
              Department of Physics, Northeast
Normal University, Changchun, Jilin 130024, P.R.China}

\date{Received: date / Accepted: date}

\maketitle

\begin{abstract}
A method of constructing $n^{2}\times n^{2}$ matrix realization of
Temperley-Lieb algebras is presented. The single loop of these
realizations are $d=\sqrt{n}$. In particular, a $9\times9-$matrix
realization with single loop $d=\sqrt{3}$ is discussed. A unitary
Yang-Baxter $\breve{R}(\theta,q_{1},q_{2})$ matrix is obtained via
the Yang-Baxterization process. The entanglement properties and
geometric properties (\emph{i.e.}, Berry Phase) of this Yang-Baxter
system are explored.
\end{abstract}

\section{Introduction}
\label{intro} Quantum Entanglement(QE)\cite{ben1,ben2,ben3,murao},
the most surprising non-classical property of a quantum system,
plays a key role in quantum information and quantum computation
processing. Similarly, topological entanglement(TE)\cite{kauffman1}
is described in terms of link diagrams and via the Artin braid
group. There are natural relationships between QE and
TE\cite{kauffman2,zhang1}. Kauffman and his co-workers have explored
the role of the unitary solutions to the Yang-Baxter
Equation(YBE)\cite{yang,baxter,drin}in quantum computation.
According to their theories, the unitary Yang-Baxter $\breve{R}$
matrices are both universal for quantum computation and are also
solutions to the condition for topological braiding. This motivates
a novel way to study YBE(as well as braid
relations)\cite{zhang2,zhang3,chen1,chen2,chen3,hu1,hu2}. A set of
size $4\times4$ universal quantum gates are constructed in terms of
unitary $\breve{R}$ matrices, for example, the CNOT
gate\cite{kauffman2}, DCNOT gate(\emph{i.e.}, Double CNOT
gate)\cite{wang1}. By means of universal $\breve{R}$ matrix,
entanglement swapping and Yang-Baxter Hamiltonian are investigated
in Ref.\cite{chen1}. In Ref.\cite{chen3}, Chen \emph{et al.} point
out that all pure two-qudit entangled states can be achieved via a
universal Yang-Baxter $\breve{R}$ matrix assisted by local unitary
transformation. Later on, the geometric properties of this
Yang-Baxter system is studied in Ref.\cite{wang2}.

 Temperley-Lieb algebras(TLA) grew out of a study of solvable lattice models in
 two-dimensional Statistical Mechanics\cite{TLA} and is related to link
and knot invariants\cite{wda}, a recent study\cite{zhang4} shows
that TLA is found to present a suitable mathematical framework for
describing quantum teleportation, entanglement swapping, universal
quantum computation and quantum computation flow. Additionally, the
systems of qutrits or more generally qudits are more powerful than
the systems of qubits habitually used in quantum
computer\cite{peres,Dkas,bruss,durt,bogda,hugh,Ospina}. Due to the
importance of TLA in quantum information processing, we find matrix
realizations of TLA in high dimension in this paper. Consequently,
by means of Yang-Baxterization approach, a family of universal
$n^{2}\times n^{2}$ $\breve{R}$ matrices associated with TLA can be
constructed.

This paper is organized as follows: In Sec.\ref{sec2}, we recall the
method of constructing matrix realizations of TLA which is given
by P.P.Kulish. Then we present a method of constructing
$n^{2}\times n^{2}$ matrix realizations of TLA with $n^{3}$
nonzero matrix elements. In Sec.\ref{sec3}, a unitary $n^{2}\times n^{2}$
Yang-Baxter $\breve{R}$ matrix is constructed via
Yang-Baxterization\cite{ge} acting on the $n^{2}\times n^{2}$ matrix
realizations of TLA. In Sec.\ref{sec4}, when $n$=3, we investigate
the entanglement properties of
$\breve{R}(\theta,q_{1},q_{2})$-matrix. We show that arbitrary
degree of entanglement for two-qutrit entangled states can be
generated via the unitary $\breve{R}(\theta,q_{1},q_{2})$-matrix
acting on the standard basis. Then we can construct a Hamiltonian
from the unitary $\breve{R}(\theta,q_{1},q_{2})$-matrix. Furthermore, the Berry
phase of the system is investigated, and the results show that the
Berry phase of this system can be interpreted under the framework of
SU(2) algebra. This result is consistent with that given in
Ref.\cite{wang2}.
\section{An extended method of constructing
realizations of TLA}\label{sec2}
In this paper,the matrix realization of TLA U-matrix and YBE
solution $\breve{R}-$matrix are $n^{2}\times n^{2}$ matrices acting on
the tensor product space $\mathcal {V} \times \mathcal {V}$, where
$\mathcal {V}$ is a $n-$dimensional vector space. As $U$ and
$\breve{R}$ act on the tensor product $\mathcal {V}_{i} \times
\mathcal {V}_{i+1}$, we denote them by $U_{i}$ and $\breve{R}_{i}$,
respectively.

 We first briefly review the theory of TLA\cite{TLA}.
For each natural number $m$, the TLA $TL_{m}(d)$ is generated by
$\{I,U_{1},U_{2}\cdots U_{m-1}\}$ with the TLA relations:
\begin{eqnarray}\label{tla}
\left\{
\begin{array}
[c]{ll} U_{i}^{2}=dU_{i} & 1\leq i \leq m-1\\
&\\
U_{i}U_{i\pm1}U_{i}=U_{i} & 1\leq i\leq m\\
& \\
U_{i}U_{j}=U_{j}U_{i} & \left\vert i-j\right\vert \geq2
\end{array}
\right.
\end{eqnarray}
 where the notation $U_{i}\equiv U_{i,i+1}$ is used. The $U_{i}$
represents $1_{1}\otimes 1_{2}\otimes 1_{3}\cdots \otimes
1_{i-1}\otimes U\otimes 1_{i+2}\otimes \cdots \otimes 1_{m}$ , and
$1_{j}$ represents the unit matrix in the $j$th space $\mathcal {V}_{j}$. In
topology, the parameter $d$ corresponds to a single loop ``$\bigcirc$''. In addition, the TLA is
easily understood in terms of knot diagrams in Ref.\cite{kauffman2}.

 In Ref.\cite{kulish}, P.P.Kulish \emph{et al.} showed a method of
constructing matrix realizations of TLA. Let us review it briefly.
For a given invertible $n\times n$ matrix A, a $n^{2}\times n^{2}$
matrix solution can be constructed in terms of A and $A^{-1}$ with
$U^{ab}_{cd}=A^{a}_{b}(A^{-1})^{c}_{d}$. Hereafter, $U^{ab}_{cd}$
denotes $U_{ab,cd}$ and $A^{a}_{b}$ denotes $A_{a,b}$ with
$a,b,c,d=0,1,2, \cdots, n-1$. One can verify that U is a matrix
realization of TLA. Let $Tr(M)$ denote the trace of matrix M, and
$M^{T}$ denote the transpose of matrix $M$. In terms of A and
$A^{-1}$, the single loop $d$ can be determined by
$d=Tr(A^{T}A^{-1})$. By means of this method, many realizations of
TLA can be constructed. For example, we set
\begin{eqnarray*}
\begin{array}{cc}
   A=\left(
                                                          \begin{array}{cc}
                                                            q^{1/2} & 0 \\
                                                            0 & q^{-1/2} \\
                                                          \end{array}
                                                        \right)
 & and ~~~ A^{-1}=\left(
                                                          \begin{array}{cc}
                                                            q^{-1/2} & 0 \\
                                                            0 & q^{1/2} \\
                                                          \end{array}
                                                        \right)
\end{array}.
\end{eqnarray*}
Then a $4\times 4$ matrix realization of TLA can be constructed as follows
\begin{eqnarray}
  U &=& \left(
           \begin{array}{cccc}
             1 & 0 & 0 & q \\
             0 & 0 & 0 & 0 \\
             0 & 0 & 0 & 0 \\
             q^{-1} & 0 & 0 & 1 \\
           \end{array}
         \right).
\end{eqnarray}
However, Not all solutions can be constructed with this method (for
example, the solution associated with eight vertex model can't be
constructed). In the following, we will introduce an extended method
of constructing matrix realizations of TLA.

In order to construct a matrix realization of TLA, we introduce two $n\times n$
invertible matrices A and B. We assume that matrix U can be
constructed as $U^{ab}_{cd}=A^{a}_{b}B^{c}_{d}$. Substituting this
relation into TLA relations (\ref{tla}), the limited conditions for
A and B can be derived. The relation $U^{2}=dU$ yields
$d=Tr(B^{T}A)$. Then U is a realization of TLA relations (\ref{tla}) if
and only if A and B respect the following conditions
\begin{equation}\label{con1}
    (BA)^{T}(AB)=(AB)(BA)^{T}=I_{n\times n}.
\end{equation}
Where $I_{n\times n}$ represents the unit matrix in $n$ dimension.
In particular, if we take $B=A^{-1}$, we note that the condition
(\ref{con1}) is obviously satisfied. Thus we re-obtain P.P.Kulish's
method of constructing matrix realizations of TLA.

In order to find matrices A and B satisfying these conditions, a
special matrix structure is adopted, that is, each row and each
column has one matrix element, and the matrix element locations are
on the main diagonal symmetry. In addition, A and B satisfy the
relation $B^{a}_{b}=(A^{a}_{b})^{-1}$ for the non-vanishing entries.
In addition, we can verify that relations
$A^{T}B=B^{T}A=AB^{T}=BA^{T}=I_{n\times n}$ hold. In this case,
Eq.(\ref{con1}) hold. Then we obtain a matrix realization of TLA.
One can verify that $Tr(I_{n\times n})=n$.

In fact, we can select $n$ matrices which satisfy these conditions,
and all their matrix elements occupy different locations. Let $i$
denote the $i$th matrix. Namely, the non-vanishing matrix elements
of $A^{(i)}$ are $(A^{(i)})^{0}_{i-1}$, $(A^{(i)})^{1}_{i-2}$,
$(A^{(i)})^{2}_{i-3}$, $\cdots$ , $(A^{(i)})^{i-1}_{0}$,
$(A^{(i)})^{i}_{n-1}$, $\cdots$ , $(A^{(i)})^{n-1}_{i}$. For
example, if $n$=4 and $i$=2, the non-vanishing matrix elements of
$A^{(2)}$ are $(A^{(2)})^{0}_{1}$, $(A^{(2)})^{1}_{0}$,
$(A^{(2)})^{2}_{3}$ and $(A^{(2)})^{3}_{2}$. There are $n$ invertible
matrix $B^{(i)}$ which is determined by
$[B^{(i)}]^{a}_{b}=[(A^{(i)})^{a}_{b}]^{-1}$ for non-varnishing
matrix elements. In terms of matrices $A^{(i)}$ and $B^{(i)}$, a
matrix representation of TLA can be constructed as
$[U^{(i)}]^{ab}_{cd}=[A^{(i)}]^{a}_{b}[B^{(i)}]^{c}_{d}$.
By means of these $n$ matrix realizations of TLA $U^{(i)}$, we can
construct a $n^{2}\times n^{2}$ matrix realization of TLA with
$n^{3}$ nonzero matrix elements. Taking the summation of these $n$ matrices, we can obtain a combined matrix
\begin{equation}\label{comb}
    U=\frac{1}{\sqrt{n}}\sum_{i=1}^{n}U^{(i)}.
\end{equation}
In fact, the realizations of TLA matrix of this form can be represented in terms of
Dirac's ``bra'' and ``ket''. This notation will appear in
Sec.\ref{sec2}. If we substitute Eq.(\ref{comb}) into
Eqs.(\ref{tla}), the first equation in Eqs.(\ref{tla}) is satisfied
automatically, and one can check that $d=\sqrt{n}$. Then the other
two relations are satisfied by the following limiting conditions,
\begin{eqnarray}\label{limited}
  \sum_{j=1}^{n}(B^{(i)}A^{(j)})^{T}(A^{(k)}B^{(j)}) &=& \textbf{0}_{n\times n}\nonumber \\
  \sum_{j=1}^{n}(A^{(j)}B^{(i)})(B^{(j)}A^{(k)})^{T} &=& \textbf{0}_{n\times
  n}.
\end{eqnarray}
Where $i\neq k$ and $i,k=1,2,\cdots n$, and $\textbf{0}_{n}$ denotes
$n\times n$ matrix with all matrix elements are zero. This limiting
condition together with the special matrix structure are used to
determine U matrix. Two examples are shown to illustrate the
application of this method in detail.

\subsection{Example I: The case $n=2$}
The simplest example which illustrates the method is the case
$n$=2. According to the above analysis, when $n=2$, we choose two
sets of $2\times2$ invertible matrices as follows,
\begin{eqnarray}
\begin{array}{cc}
  A^{(1)}=\left(
          \begin{array}{cc}
            a_{1} & 0 \\
            0 & b_{1} \\
          \end{array}
        \right),
   & B^{(1)}=\left(
          \begin{array}{cc}
            a_{1}^{-1} & 0 \\
            0 & b_{1}^{-1} \\
          \end{array}
        \right) \\
        \\
  A^{(2)}=\left(
          \begin{array}{cc}
            0 & a_{2} \\
            b_{2} & 0 \\
          \end{array}
        \right), & B^{(2)}=\left(
          \begin{array}{cc}
            0 & a_{2}^{-1} \\
            b_{2}^{-1} & 0 \\
          \end{array}
        \right).
\end{array}
\end{eqnarray}
Where $a_{i}$ and $b_{i}$ are parameters which will be determined by
the conditions in Eq.(\ref{limited}). Then two U matrices can be
obtained as follows (we choose $\{|00\rangle, |01\rangle,
|10\rangle, |11\rangle\}$ as standard basis),
\begin{eqnarray}
\begin{array}{cc}
  U^{(1)}=\left(
          \begin{array}{cccc}
            1 & 0 & 0 & a_{1}b_{1}^{-1} \\
            0 & 0 & 0 & 0 \\
            0 & 0 & 0 & 0 \\
            a^{-1}_{1}b_{1} & 0 & 0 & 1 \\
          \end{array}
        \right),
   &  U^{(2)}=\left(
          \begin{array}{cccc}
            0 & 0 & 0 & 0 \\
            0 & 1 & a_{2}b_{2}^{-1} & 0 \\
            0 & a^{-1}_{2}b_{2} & 1 & 0 \\
            0 & 0 & 0 & 0 \\
          \end{array}
        \right)
\end{array}.
\end{eqnarray}
The trace of these two solutions is $2$ (\emph{i.e.},
$d_{1}=d_{2}=2$).

In order to obtain a solution associated with eight-vertex model, we
consider the combinatorial structure of $U^{(1)}$ and $U^{(2)}$. The
combinatorial form reads $$U=\frac{1}{\sqrt{2}}(U^{(1)}+U^{(2)}).$$ If we substitute this relation into
Eqs.(\ref{limited}). Then we can derive a strong limiting condition
$a_{2}b_{2}^{-1}=\epsilon i$
($\epsilon=\pm$). Let $M^{*}$ denote complex conjugation of matrix $M$. We can
introduce a new parameter $q$ with $q=a_{1}b_{1}^{-1}$, which is
complex and has norm 1(\emph{i.e.} $q^{*}=q^{-1}$). Then a
eight-vertex matrix representation with $d=\sqrt{2}$ is obtained as
follows,
\begin{eqnarray}\label{2}
  U &=& \frac{1}{\sqrt{2}}\left(
           \begin{array}{cccc}
             1 & 0 & 0 & q \\
             0 & 1 & \epsilon i & 0 \\
             0 & -\epsilon i & 1 & 0 \\
             q^{-1} & 0 & 0 & 1 \\
           \end{array}
         \right).
\end{eqnarray}
Let $$|\psi_{1}\rangle=\frac{1}{\sqrt{2}}(|00\rangle+q^{-1}|11\rangle),$$
$$|\psi_{2}\rangle=\frac{1}{\sqrt{2}}(|01\rangle-\epsilon i|10\rangle).$$
Then, in terms of ``bra'' and ``ket'', the U matrix takes the following form
$$U=\sqrt{2}(|\psi_{1}\rangle\langle\psi_{1}|+|\psi_{2}\rangle\langle\psi_{2}|).$$
This realization of TLA is
associated with eight vertex model\cite{baxter}. And this solution
has been applied to many fields, such as topological quantum
computation\cite{nayak} and two dimensional representation of
YBE\cite{hu1}.
\subsection{Example II: The case $n=3$}
Let $A^{(i)}$ and $B^{(i)}$ (i=1,2,3) are three sets of $3\times 3$
matrices with standard basis(\emph{i.e.}, $|0\rangle, |1\rangle,
|2\rangle$). We set

\begin{eqnarray}
\begin{array}{cc}
  A^{(1)}=\left(
          \begin{array}{ccc}
            0 & 0 & a_{1} \\
            0 & b_{1} & 0 \\
            c_{1} & 0 & 0 \\
          \end{array}
        \right)
   & B^{(1)}=\left(
          \begin{array}{ccc}
            0 & 0 & a_{1}^{-1} \\
            0 & b_{1}^{-1} & 0 \\
            c_{1}^{-1} & 0 & 0 \\
          \end{array}
        \right) \\
        \\
  A^{(2)}=\left(
          \begin{array}{ccc}
            0 & a_{2} & 0 \\
            b_{2} & 0 & 0 \\
            0 & 0 & c_{2} \\
          \end{array}
        \right)
   & B^{(2)}=\left(
          \begin{array}{ccc}
            0 & a_{2}^{-1} & 0 \\
            b_{2}^{-1} & 0 & 0 \\
            0 & 0 & c_{2}^{-1} \\
          \end{array}
        \right) \\
        \\
  A^{(3)}=\left(
          \begin{array}{ccc}
            a_{3} & 0 & 0 \\
            0 & 0 & b_{3} \\
            0 & c_{3} & 0 \\
          \end{array}
        \right)
   & B^{(3)}=\left(
          \begin{array}{ccc}
            a_{3}^{-1} & 0 & 0 \\
            0 & 0 & b_{3}^{-1} \\
            0 & c_{3}^{-1} & 0 \\
          \end{array}
        \right).
\end{array}
\end{eqnarray}
Where $a_{i}$, $b_{i}$ and $c_{i}$ are also undetermined parameters. Thus we note that the relation Eq.(\ref{con1})is clearly
satisfied. If we choose $\{|00\rangle,|01\rangle,|02\rangle,|10\rangle,|11\rangle,|12%
\rangle,|20\rangle,|21\rangle,|22\rangle\}$ as standard basis, then
we can obtain three sets of $3^{2}\times 3^{2}$ matrices $U^{(1)}$,
$U^{(2)}$ and $U^{(3)}$. In this case, their single loop
$d_{i}=3$($i$=1, 2, 3). Then the combined form of U matrix
$U=(U^{(1)}+U^{(2)}+U^{(3)})/\sqrt{3}$. Substituting this combined
form into Eqs.(\ref{limited}), the undetermined parameters follows
from the limited conditions,
\begin{eqnarray}
  \begin{array}{ccc}
    a_{1}b_{1}^{-1}=\frac{q_{1}}{q_{2}} & a_{2}b_{2}^{-1}=\omega &
    a_{3}b_{3}^{-1}=\omega q_{1}\\
    \\
 a_{1}c_{1}^{-1}=1, & a_{2}c_{2}^{-1}=\omega q_{2}, &
    a_{3}c_{3}^{-1}=q_{1}.
  \end{array}
\end{eqnarray}
Where $q_{i}=e^{i\varphi_{i}}$ and $\omega$ satisfies the relation
$\omega^{2}+\omega+1=0$ (\emph{i.e.}, $\omega=e^{i\epsilon
\frac{2\pi}{3}}$). On the standard basis U has the matrix form
\begin{eqnarray}\label{3}
 U=\frac{1}{\sqrt{3}}\left(
  \begin{array}{ccccccccc}
    1 & 0 & 0 & 0 & 0 & \omega q_{1} & 0 & q_{1} & 0 \\
    0 & 1 & 0 & \omega & 0 & 0 & 0 & 0 & \omega q_{2} \\
    0 & 0 & 1 & 0 & \frac{q_{1}}{q_{2}} & 0 & 1 & 0 & 0 \\
    0 & \frac{1}{\omega} & 0 & 1 & 0 & 0 & 0 & 0 & q_{2} \\
    0 & 0 & \frac{q_{2}}{q_{1}} & 0 & 1 & 0 & \frac{q_{2}}{q_{1}} & 0 & 0 \\
    \frac{1}{\omega q_{1}} & 0 & 0 & 0 & 0 & 1 & 0 & \frac{1}{\omega} & 0 \\
    0 & 0 & 1 & 0 & \frac{q_{1}}{q_{2}} & 0 & 1 & 0 & 0 \\
    \frac{1}{q_{1}} & 0 & 0 & 0 & 0 & \omega & 0 & 1 & 0 \\
    0 & \frac{1}{\omega q_{2}} & 0 & \frac{1}{q_{2}} & 0 & 0 & 0 & 0 & 1 \\
  \end{array}
\right).
\end{eqnarray}
The single loop of this solution is $d=\sqrt{3}$. In fact, we can
introduce three sets maximally entangled states as
\[|\psi_{1}\rangle=\frac{1}{\sqrt{3}}(|02\rangle+q_{1}q_{2}^{-1}|11\rangle+|20\rangle)\]
$$|\psi_{2}\rangle=\frac{1}{\sqrt{3}}(|01\rangle+\omega^{-1}|10\rangle+\omega^{-1}
q^{-1}_{2}|22\rangle)$$
$$|\psi_{3}\rangle=\frac{1}{\sqrt{3}}(|00\rangle+\omega^{-1}
q_{1}^{-1}|12\rangle+q_{1}^{-1}|21\rangle).$$ In terms of
these maximally entangled states, the U matrix (\ref{3}) can be
written in a elegant form
$$U=\sqrt{3}(|\psi_{1}\rangle\langle\psi_{1}|+|\psi_{2}\rangle\langle\psi_{2}|+|\psi_{3}\rangle\langle\psi_{3}|).$$
\subsection{Remarks}
We close this section with some remarks. Juan Ospina made a
Mathematica implementation of this method, and the results in this
article were re-obtained\cite{ospina}. When $n=2$, the solution
(\ref{2}) has been discussed in many works. As we all know, when
$n=3$, the solution(\ref{3}) is not discussed. We note that the
solutions(\ref{2}) and (\ref{3}) are Hermitian matrices
(\emph{i.e.}, $U^{\dag}=U$)(This fact will be used in the process of
Yang-Baxterization approach).
\section{Yang-Baxterization of U matrix}\label{sec3}
In order to discuss the non-maximally entangled states, the author
In Ref.\cite{chen1}, the unitary $\breve{R}$ matrix has been
introduced in Ref.\cite{chen1}. To make the paper self-contained, we
briefly review it in the following. In this work, we utilize the so
called relativistic Yang-Baxter Equation(YBE)\cite{hu1}. The
relativistic YBE reads,
\begin{eqnarray}\label{ybe}
\breve{R}_{i}(u)\breve{R}_{i+1}\left(\frac{u+v}{1+ \beta^2
uv}\right)\breve{R}_{i}(v)=
\breve{R}_{i+1}(v)\breve{R}_{i}\left(\frac{u+v}{1+ \beta^2
uv}\right)\breve{R}_{i+1}(u)
\end{eqnarray}
where $\breve{R}_{i}$ represents $1_{1}\otimes 1_{2}\otimes
1_{3}\otimes \cdots \otimes 1_{i-1}\otimes \breve{R}\otimes
1_{i+2}\otimes \cdots \otimes 1_{m}$. The variables $u$ and $v$ are called as
the spectral parameters. The $\beta$ is a constant with $\beta^{-1}=i
c$ ($c$ is light velocity).

Let the unitary $\breve{R}(u)$ matrix take the form
\begin{equation}\label{ybz}
    \breve{R}(u)=F(u)[I_{n\times n}+G(u)U].
\end{equation}
Where the functions $F(u)$ and $G(u)$ are to be determined.
Substituting Eq.(\ref{ybz}) into Eq.(\ref{ybe}), we obtain the
relation
\begin{equation}\label{yangbaxter}
    G(u)+G(v)+G\left(\frac{u+v}{1+\beta^{2}uv}\right)[G(u)G(v)-1]+\sqrt{n}G(u)G(v)=0
\end{equation}
Following Hu \emph{et al.}\cite{hu1}, we set $$G(u)=\frac{a\beta
u}{b\beta^{2}u^{2}+c\beta u+d}.$$ If we substitute it to the
relation Eq.(\ref{yangbaxter}). Then we obtain equations for undetermined parameters $a, b ,c$ and $d$,
\begin{eqnarray*}
\left\{
\begin{array}
[c]{ll} a^{2}+\sqrt{n}ac+c^{2}+3bd+d^{2}=0 &\\
&\\
\sqrt{n}a+2c=0 & \\
& \\
b=d &
\end{array}.
\right.
\end{eqnarray*}
After some algebra, a solution of $G(u)$ is obtained as follows

\begin{equation}\label{gu}
    G(u)=\frac{4i\epsilon\beta u}{\sqrt{4-n}(\beta^{2}u^{2}-2\sqrt{n/(4-n)}i\epsilon\beta
    u+1)}.
\end{equation}
We note that $n\neq 4$. The case $d=\sqrt{2}$ has been discussed in Ref.\cite{hu1}. In addition,
the unitary relation
$\breve{R}^{\dag}(u)\breve{R}(u)=\breve{R}(u)\breve{R}^{\dag}(u)=I_{n\times
n}$ leads to the relation $F^{*}(u)F(u)=1$ and
$G(u)+G^{*}(u)+\sqrt{n}G(u)G^{*}(u)=0$, where $*$ denotes complex
conjugation. Consider these relations, one can introduce a new
variable $\theta$ with $G(u)=(e^{-2i\theta}-1)/\sqrt{n}$, which
equivalent to the relation
$$\frac{\beta^{2}u^{2}+2\sqrt{n/(4-n)}i\epsilon\beta
u+1}{\beta^{2}u^{2}-2\sqrt{n/(4-n)}i\epsilon\beta
u+1}=e^{-2i\theta}.$$ We set $F(u)=e^{i\theta}$ with $\theta$ is
real. In terms of the new variable, we rewrite the Yang-Baxter
matrix in a new form
\begin{equation}\label{ybe2}
   \breve{R}(\theta,q_{1},q_{2})=e^{i\theta}I_{n\times n}-\frac{2isin\theta}{\sqrt{n}}U.
\end{equation}
The case of $n$=2 has been discussed in Ref.(\cite{hu1}). If $n$=3,
on the standard basis the unitary solution of $\breve{R}$ matrix is
\begin{eqnarray}\label{rmatrix}
 \breve{R}=\frac{1}{3}\left(
  \begin{array}{ccccccccc}
    f & 0 & 0 & 0 & 0 & \omega g q_{1} & 0 & gq_{1} & 0 \\
    0 & f & 0 & \omega g & 0 & 0 & 0 & 0 & \omega g q_{2} \\
    0 & 0 & f & 0 & g\frac{q_{1}}{q_{2}} & 0 & g & 0 & 0 \\
    0 & \frac{g}{\omega} & 0 & f & 0 & 0 & 0 & 0 & gq_{2} \\
    0 & 0 & g\frac{q_{2}}{q_{1}} & 0 & f & 0 & g\frac{q_{2}}{q_{1}} & 0 & 0 \\
    \frac{g}{\omega q_{1}} & 0 & 0 & 0 & 0 & f & 0 & \frac{g}{\omega} & 0 \\
    0 & 0 & g & 0 & g\frac{q_{1}}{q_{2}} & 0 & f & 0 & 0 \\
    \frac{g}{q_{1}} & 0 & 0 & 0 & 0 & \omega g & 0 & f & 0 \\
    0 & \frac{g}{\omega q_{2}} & 0 & \frac{g}{q_{2}} & 0 & 0 & 0 & 0 & f \\
  \end{array}
\right).
\end{eqnarray}
Where $f\equiv f(\theta)=(e^{-i\theta}+2e^{i\theta})/\sqrt{3}$ and
$g\equiv g(\theta)=(e^{-i\theta}-e^{i\theta})/\sqrt{3}$.
\section{Entanglement and Hamiltonian}\label{sec4}

By Brylinski＊s theorem\cite{brylinski}, a $4\times4$ Yang-Baxter
$\breve{R}$ matrix is universal for quantum computation, if and only
if this Yang-Baxter $\breve{R}$ matrix can generate entangled states
from separable states. The proof of universality for $n^{2}\times
n^{2}$ Yang-Baxter matrix is presented in Ref.\cite{chen3}. Via a
unitary universal Yang-Baxter $\breve{R}$ matrix acting on the
standard basis, one can obtain a set of entangled states. For
example, if one lets $\breve{R}(\theta)$ act on the separable state
$|lm\rangle$(\emph{i.e.}, $|l\rangle\otimes|m\rangle$), this yields
the following family of states
$|\psi\rangle_{lm}=\sum_{ij=00}^{n-1,n-1}\breve{R}^{ij}_{lm}|lm\rangle$($l,m=0,1,\cdots
, n-1$). These unitary matrices may be universal for quantum
computation, hence they can entangle states. The case $n$=2 has been
discussed in Ref.\cite{chen1}.

Hereafter we focus on the case $n$=3. For example, if $l$=0 and
$m$=0, then
$|\psi\rangle_{00}=(f|00\rangle+\omega^{-1}gq_{1}^{-1}|12\rangle+
gq_{1}^{-1}|21\rangle)/3$.  By means of negativity, we study these
entangled states. It should be noted that the negativity criterion
is necessary and sufficient only for $2\otimes 2$ and $2 \otimes 3$
quantum systems. However, negativity is well-defined for
calculation, and it has been widely applied to evaluation of
entanglement \cite{zycz,wangxg,RYR}. The negativity criterion for
two qutrits is given by
\begin{equation}
\mathcal{N}(\rho)\equiv\frac{\|\rho^{T_{A}}\|_{1}-1}{2},
\end{equation}
where $\|\rho^{T_{A}}\|_{1}$ denotes the trace norm of
$\rho^{T_{A}}$, $\rho^{T_{A}}$ denotes the partial transpose of the
bipartite state $\rho$. The $\mathcal{N}(\rho)$ corresponds to the
absolute value of the sum of negative eigenvalues of $\rho^{T_{A}}$,
and negativity vanishes for unentangled states. Then negativity of
the state $|\psi\rangle_{00}$ yields
\begin{equation}\label{N}
\mathcal{N}(\theta)=\frac{4}{9}(sin^{2}\theta+|\sin\theta|\sqrt{%
1+8cos^{2}\theta}).
\end{equation}
If $|g|=|f|$(\emph{i.e.} $x=e^{i\frac{\pi}{3}}$), then the state
$|\psi\rangle_{00}$ becomes the maximally entangled state for two
qutrits $$|\psi\rangle_{00}=%
\frac{1}{\sqrt{3}}(e^{i\frac{\pi}{6}}|00\rangle-i\omega^{-1}q_{1}^{-1}|12\rangle-iq_{1}^{-1}|21\rangle).$$
In general, the unitary Yang-Baxter matrix $\breve{R}(\theta)$ acts
on the
basis $\{|00\rangle$, $|01\rangle$, $|02\rangle$, $|10\rangle$, $|11\rangle$, $|12%
\rangle$, $|20\rangle$, $|21\rangle$, $|22\rangle\}$, we obtain the same
range of negativity as Eq(\ref{N}). It is easy to check that the
negativity ranges from 0 to 1 when the parameter $\theta$ runs from
0 to $\pi$. But for $\theta \in [0,\pi]$, the negativity is not a
monotonic function of $\theta$. And when $\theta=\pi/3$,
$\breve{R}(\theta)$ generate nine complete and orthogonal maximally
entangled states for two qutrits.

In fact, we can introduce a unitary transformation $Y=Y_{1}\otimes
Y_{2}$. $Y_{1}$ and $Y_{2}$ take the form

\begin{eqnarray*}
 \begin{array}{cc}
  Y_{1}=\left(
                       \begin{array}{ccc}
                         e^{i\frac{4\pi}{9}} & 0 & 0 \\
                         0 & 1 & 0 \\
                         0 & 0 & e^{-i\frac{4\pi}{9}} \\
                       \end{array}
                     \right) & Y_{2}=\left(
                                           \begin{array}{ccc}
                                             e^{-i\frac{2\pi}{9}} & 0 & 0 \\
                                             0 & 1 & 0 \\
                                             0 & 0 & e^{-i\frac{4\pi}{9}} \\
                                           \end{array}
                                         \right).
 \end{array}
\end{eqnarray*}
By means of this local transformation, the universal $\breve{R}$
matrix(\ref{rmatrix}) is locally equivalent to $\breve{R}$ matrix in
Ref.\cite{wang2}.

A Hamiltonian of the Yang-Baxter system can be constructed from the
$\breve{R}(\theta , \varphi_{1},\varphi_{2})$-matrix. As shown in
Ref.\cite{hu1}, the Hamiltonian is obtained through the
Schr\"{o}dinger evolution of the entangled states. Let the
parameters $\varphi_{i}$ be time-dependent as
$\varphi_{i}=\omega_{i} t$. The Hamiltonian is
\begin{eqnarray}
\hat{H} &=& i\hbar\frac{\partial\breve{R}(\theta ,
\varphi_{1},\varphi_{2})}{\partial t}\breve{R}^{\dag }(\theta ,
\varphi_{1},\varphi_{2}).
\end{eqnarray}
This Hamiltonian is equivalent to the Hamiltonian in
Ref.\cite{wang2}, so one can obtain the same results as in
Ref.\cite{wang2}. The Berry phase of this system can also be
explained in the framework of SU(2) algebra. The Berry phase can be
explained as solid angle which is expanded in the parameter space.
We will not discuss this in detail in this paper. But we should note
that the meaning of the parameter $\theta$ is different. The
$\theta$ in Ref.(\cite{wang2}) arises from trigonometrical
parameterization, and the $\theta$ in this work arises from the
relativistic rational parameter.

\section{Summary}
In this paper, we present a method of constructing $n^{2}\times
n^{2}$ matrix realization of TLA. This matrix realization of
TLA has $n^{3}$ nonzero matrix elements. Applying Yang-Baxterization
approach to the matrix realization of TLA, one can obtain a
$n^{2}\times n^{2}$ Yang-Baxter $\breve{R}$ matrix. When a
Yang-Baxter $\breve{R}$ matrix acts on the standard basis, one can
obtain a family of entangled states. Yang-Baxter $\breve{R}$ matrix
is universal for quantum computation.

We believe that this family of Yang-Baxter $\breve{R}$ matrices
associated with U matrices will be applied in quantum information, quantum
computation and so on. We will investigate these applications in
subsequent papers.

\begin{acknowledgements}
The authors gratefully acknowledge Juan Ospina for helpful comments
on this paper. Special thanks to the first referee for his advice
and criticism on our manuscript.
\end{acknowledgements}


\begin{thebibliography}{}
\bibitem{ben1} C. H. Bennett and D. P. DiVincenzo.:Quantum information and computation. Nature \textbf{404}
247(2000).
\bibitem{ben2} C. H. Bennett and G. Brassard, C. Cr\'{e}peau, R. Jozsa, A Peres, and W. K. Wootters.:Teleporting
an unknown quantum state via dual classical and
Einstein-Podolsky-Rosen channels. Phys. Rev. Lett. \textbf{70},
1895(1993).
\bibitem{ben3} C H. Bennett and S. J. Wiesner.:Communication via one- and two-particle operators on
Einstein-Podolsky-Rosen states. Phys. Rev. Lett. \textbf{69},
2881(1992).
\bibitem{murao} M. Murao, D. Jonathan, M. B. Plenio, and V. Vedral.:Quantum telecloning and multiparticle
entanglement, Phys. Rev. \textbf{A 59}, 156(1999).
\bibitem{kauffman1}L. H. Kauffman.: Knots and Physics, World Scientific Publishers(2002).
\bibitem{kauffman2}L. H. Kauffman and S. J. Lomonaco Jr.:Braiding operators are universal quantum gates. New J. Phys.\textbf{4},73.1每73.18.(2002).
\bibitem{zhang1} Yong Zhang,Louis H. Kauffman, and Mo-Lin Ge.:Yang每Baxterizations, Universal Quantum Gates and Hamiltonians, Quantum Information Processing, Vol. \textbf{4}, No. 3, August
(2005).
\bibitem{yang} C. N. Yang.: Some Exact results for the many-body problem in one dimension with
repulsive delta-function interaction. Phys. Rev. Lett. \textbf{19},
1312(1967); C. N. Yang.: S matrix for the one-dimensional N-body
problem with repulsive or attractive -function interaction. Phys.
Rev. \textbf{168} 1920(1968).
\bibitem{baxter} R. J. Baxter.:Exactly Solved Models in Statistical Mechanics Academic
Press, London, (1982); R. J. Baxter.:Partition funtion of the
eighy-vertex lattice model. Ann. Phys. \textbf{70}, 193(1972).
\bibitem{drin} V. G. Drinfeld.: Hopf algebras and the quantum Yang-Baxter equation. Soviet Math.  Dokl \textbf{32},pp. 254-258(1985).
\bibitem{zhang2} Y. Zhang, L. H. Kauffman, and M. L. Ge.: Universal quantum gate, YangBaxterization
and Hamiltonian. Int. J. Quant. Inf.\textbf{3} 669(2005).
\bibitem{zhang3} Y Zhang and M. L. Ge.: GHZ states, almost-complex structure and Yang-Baxter equation, Quant. Inf. Proc. \textbf{6} 363(2007);
Y. Zhang, E. C. Rowell, Y. S. Wu, Z. H. Wang, M. L. and Ge.: From
extraspecial two-Groups to GHZ states, e-print
quant-ph/0706.1761(2007).
\bibitem{chen1} J. L. Chen, K. Xue, and M. L. Ge.: Braiding transformation, entanglement swapping, and
Berry phase in entanglement space. Phys. Rev. A. \textbf{76},
042324(2007).
\bibitem{chen2} J. L. Chen, K. Xue, and M. L. Ge.: Berry phase and quantum criticality in Yang Baxter
systems. Ann. Phys. \textbf{323} 2614(2008).
\bibitem{chen3} J. L. Chen, K. Xue, and M. L. Ge.: All pure two-qudit entangled states can be generated via
a universal Yang每Baxter matrix assisted by local unitary
transformations. Chinese Phys. Lett. \textbf{26}, 080306 (2009).
\bibitem{hu1}Shuang-Wei Hu,Kang Xue, and Mo-Lin Ge.: Optical simulation of the Yang-Baxter equationPhys. Rev. A
\textbf{78}, 022319(2008).
\bibitem{hu2} Ming-Guang Hu,Kang Xue, and Mo-Lin Ge.: Exact Solution of a Yang-Baxter Spin-1/2 Chain Model and Quantum Entanglement. Phys. Rev. A \textbf{78}, 052324 (2008)
\bibitem{wang1} Gangcheng Wang, Kang Xue, Chunfeng Wu, He Liang and C H
Oh.: Entanglement and the Berry phase in a new Yang-Baxter system.
J. Phys. A: Math. Theor. \textbf{42}, 125207(2009).
\bibitem{wang2} Gangcheng Wang, Chunfang Sun, Qingyong Wang, kang Xue, Entanglement and Berry Phase in a $(3\times3)$-dimensional Yang-Baxter System , International Journal of Theoretical
Physics, doi: 10.1007/s10773-009-0077-z
\bibitem{TLA}H. Temperley, E.H. Lieb.: Relations between the ※percolation§ and ※colouring§ problem and other graph-theoretical problems associated with regular planar lattices: some exact results for the ※percolation§ problem,  Proc. Roy. Soc. (London)
\textbf{A322}, 251 (1971)
\bibitem{wda}Wadati M, Deguchi T and
Akutsu Y.: Exactly solvable models and knot theory, Phys.
Rep.\textbf{180}, 247(1989)
\bibitem{zhang4}Yong Zhang.: Teleportation, braid group and Temperley--Lieb algebra. J. Phys. A: Math. Gen.
\textbf{39}, 11599-11622(2006)
\bibitem{peres}H. Bechmann-Pasquinucci and A. Peres.: Quantum Cryptography with 3-State Systems. Phys. Rev. Lett.
\textbf{85}, 3313 (2000).
\bibitem{Dkas}D. Kaszlikowski \emph{et al.}.: Quantum cryptography based on qutrit Bell inequalities.  Phys. Rev. A \textbf{67}, 012310 (2003).
\bibitem{bruss}D. Bruss and C. Machiavello.: Optimal Eavesdropping in Cryptography with Three-Dimensional Quantum States
. Phys. Rev. Lett. \textbf{88}, 127901 (2002).
\bibitem{durt}Yong-Cheng Ou \emph{et al.}.: Proper monogamy inequality for arbitrary pure quantum states. Phys. Rev. A \textbf{67}, 012311 (2003).
\bibitem{bogda}Y.I. Bogdanov \emph{et al.}.: Quantum State Engineering with qutrits
. Phys. Rev. L \textbf{93}, 230503 (2004).
\bibitem{hugh}D.M.Hugh, J.Twamley.: Trapped-ion qutrit spin molecule quantum computer. New J. Phys. \textbf{7}, 174(2005) .
\bibitem{Ospina} B. Beliczynski \emph{et al.} (Eds.): ICANNGA 2007, Part I, LNCS 4431, pp. 120-127(2007).
\bibitem{ge} M.L. Ge, K. Xue and Y-S. Wu.: Explicit
Trigonometric Yang-Baxterization. Int. J. Mod. Phys. \textbf{A6},
3735(1991);
\bibitem{kulish}P. P. Kulish, N. Manojlovic, and Z. Nagy.:  Quantum symmetry algebras of spin systems related to Temperley-Lieb R-matrices. J. Math. Phys. \textbf{49}, 023510 (2008);
\bibitem{nayak} Chetan Nayak, Steven H. Simon, Ady Stern, Michael
Freedman, Sankar Das Sarma.:Non-Abelian Anyons and Topological
Quantum Computation. Rev. Mod. Phys. \textbf{80}, 1083 (2008);
\bibitem{ospina} Juan Ospina made a Mathematica implementation of
our method, and the results in this paper were re-obtained.
\bibitem{brylinski}J. L. Brylinski and R. Brylinski, Universal Quantum Gates, in R. Brylinski and G.
Chen (eds.).: Mathematics of Quantum Computation, Chapman and
Hall/CRC Press, Boca Raton, Florida(2002).
\bibitem{zycz}K. Zyczkowski, \emph{et al.}: Volume of the set of separable states. Phys. Rev. A, \textbf{58}, 883, (1998).
\bibitem{wangxg}Xiaoguang Wang \emph{et al.}.: Negativity, entanglement witnesses and quantum phase transition in spin-1 Heisenberg chains.  J. Phys. A: Math. Theor. \textbf{40}
10759-10767(2007)
\bibitem{RYR}Recep Eryigit, Yigit Guc, Resul Eryigit.:Analytical Study of Thermal Entanglment in a Two-dimensional $J_{1}-J_{2}$ model. Phys. Lett. A \textbf{358} 363, (2006).
\bibitem{Ma}Xiao San Ma.: Thermal entanglement of a two-qutrit XX spin chain with Dzialoshinski Moriya interaction. Optics Communications \textbf{281},  484每488(2008).
\end{thebibliography}


\end{document}